# Adversarial Training for Dynamics Matching in Coarse-Grained Models

Yihang Wang and Gregory A. Voth[*]

[1]Department of Chemistry, Chicago Center for Theoretical Chemistry, Institute for Biophysical Dynamics, and James Franck Institute, The University of Chicago, Chicago, IL 60637

**Corresponding Author**:

*Gregory A. Voth
Department of Chemistry
The University of Chicago
5735 S. Ellis Ave, SCL 123
Chicago, IL 60637

Phone: (773) 702-9092
Fax: (773) 795-9106
**E-mail**: gavoth@uchicago.edu

**Abstract:**

Molecular dynamics (MD) simulations are essential for studying complex molecular systems, but their high computational cost limits scalability. Coarse-grained (CG) models reduce this cost by simplifying the system, yet traditional approaches often fail to maintain dynamic consistency, compromising their reliability in kinetics-driven processes. Here, we introduce an adversarial training framework that aligns CG trajectory ensembles with all-atom (AA) reference dynamics, ensuring both thermodynamic and kinetic fidelity. Our method adapts the adversarial learning paradigm, combining a physics-based generator with a neural network discriminator that differentiates between AA and CG trajectories. By adversarially optimizing CG parameters, our approach eliminates the need for predefined kinetic features. Applied to liquid water, it accurately reproduces radial and angular distribution functions as well as dynamical mean squared displacement, even extrapolating long-timescale dynamics from short training trajectories. This framework offers a new approach for bottom-up CG modeling, offering a systematic and principled way to preserve dynamic consistency in complex coarse-grained molecular systems.

## I. Introduction

Molecular dynamics (MD) simulations have become a crucial tool in studying the behavior of complex molecular systems. However, the computational cost associated with all-atom (AA) simulations severely limits their ability to capture long timescales and large system sizes.[1] To overcome these limitations, coarse-graining (CG) techniques have been widely adopted, systematically reducing the degrees of freedom while aiming to retain the essential properties of the system of interest. Bottom-up CG modeling, which derives effective interactions by learning from high-resolution AA simulation, has proven successful in applications ranging from lipid membrane to polymer self-assembly.[2-6] Despite its success, a key challenge in any CG modeling remains: i.e., ensuring dynamic consistency, such that the CG model faithfully reproduces not only equilibrium properties but also the kinetics of the underlying atomistic system.[4, 5, 7] This limitation hinders the application of CG methods to large-scale biomolecular processes where kinetics plays a critical role, such as the aggregation of misfolded proteins.[8]

From the perspective of bottom-up CG modeling, projecting the AA dynamics into some reduced degrees of freedom (collective variables) brings more complicated interplay between these variables, raising the fundamental question of how to define the governing equation of motion for these variables. Several theoretical frameworks have been developed to answer this question, including the Mori-Zwanzig formalism,[9, 10] which formulates projected dynamics with the generalized Langevin equation (GLE) form, and the effective dynamical formula,[11] which describes the dynamics of collective variables when they sufficiently capture essential reaction mechanisms. While these frameworks elegantly formalize reduced dynamics, their practical implementation often faces challenges due to the need to compute intractable terms such as memory kernels or high-dimensional free energy landscapes.

Building upon these theoretical foundations, numerical methods have been developed to construct dynamically consistent CG models. For example, an approach based on consistent Markov State Models has been developed in an attempt to enhance the kinetic fidelity of models.[12, 13] In the work by Izvekov and Voth, friction coefficients were derived by approximating the thermal friction and interactions arising from eliminated degrees of freedom.[4] Expanding on this concept, Davtyan et al. introduced the dynamic force matching technique, which incorporates fictitious particles to capture both short- and long-term dynamical properties.[14, 15] A more recent study by Xie and E parameterized the memory kernel via a data-driven Mori-Zwanzig framework, leveraging a variational principle to systematically extract non-Markovian effects directly from AA trajectories.[16]

It should be noted that the development of machine learning (ML) models also enables the direct learning of the dynamics of partially observed molecular systems.[17-19] Compared to machine learning emulators, which directly learn surrogate models for molecular dynamics, physics-based CG models have intrinsic advantages: they enforce physical constraints and generalize better across conditions outside the training data [20]. Therefore, it is valuable to investigate how to ensure that a CG model generates realistic dynamical trajectories .

Recent advances in generative artificial intelligence (AI) offer a promising route for overcoming the challenge of building dynamically consistent CG models. In particular, adversarial training, a framework widely used in generative modeling, has demonstrated success in learning high-dimensional probability distributions.[21, 22] The fundamental idea behind adversarial learning is to iteratively improve a generator by using a discriminator that distinguishes between generated and reference samples. This principle naturally extends to CG modeling, where the goal is to match the trajectory ensemble from a generator (here, a CG model) to that of the AA model. Inspired by the adversarial-residual-coarse-graining (ARCG) approach introduced by Durumeric and Voth,[23] we apply an adversarial learning framework to dynamically match CG trajectories with their AA counterparts, leveraging neural networks as discriminators to optimize the CG parameters.

In this work, we introduce an adversarial training strategy to construct CG models with dynamic consistency across regimes such as Brownian or Langevin dynamics. Unlike traditional methods that rely on manually selected kinetic features, our approach directly matches entire trajectory ensembles. To enhance computational efficiency, we further investigate the choice of maximum stable time step for simulations, balancing accuracy with computational cost.

The remainder of this paper is structured as follows. Section II provides the theoretical background of the adversarial training framework and outlines the methodological implementation, including the neural network architectures used for the discriminator and generator. Section III presents the application of our approach to a model molecular system and evaluates its performance in recovering both equilibrium and dynamical properties. Finally, Section IV discusses broader implications, potential extensions to more complex systems, and the role of this framework in parameterizing position-dependent diffusion coefficients and memory kernels.

## II. Theory and Methods

In bottom-up CG modeling, the thermodynamic consistency[3] is described as the requirement for the distribution of CG conformations to reproduce the marginal equilibrium distribution of the all-atom system when projected onto the coarse-grained representation. That is:

$$P^{CG}(\boldsymbol{R}) = \int \delta(\mathcal{M}(\boldsymbol{r}) - \boldsymbol{R})P^{AA}(\boldsymbol{r})d\boldsymbol{r}, \qquad (1)$$

where $\boldsymbol{R}$ represents the CG coordinates, $\mathcal{M}(\boldsymbol{r})$ is the mapping operator that projects AA configurations onto the CG space, and $P^{AA}(r)$ is the equilibrium distribution of the AA system. Since the equilibrium distribution of the CG system is connected to the CG potential through relationship $P^{CG}(\boldsymbol{R}) \propto e^{-\beta U(\boldsymbol{R})}$, where $\beta = \frac{1}{k_B T}$, one can achieve thermodynamic consistency by optimizing the CG potential $U(\boldsymbol{R})$. Methods like force matching,[2, 3, 24] and relative energy minimization (REM)[19, 25, 26] have been developed as the prototype methods for this aim.

While thermodynamic consistency ensures the model preserves the statistical mechanics of the reference atomistic system, such as free energy landscapes and ensemble-averaged observables, it does not ensure the accurate reproduction of dynamical properties.[4] To address this, we introduce the principle of dynamic consistency, requiring the CG model to preserve the temporal evolution of the AA system. This is defined as:

$$P^{CG}(\boldsymbol{R_{1:T}}) = \int \prod_{t=1}^{T} \delta(\mathcal{M}(\boldsymbol{r_t}) - \boldsymbol{R_t})\, P^{AA}(\boldsymbol{r_{1:T}})d\boldsymbol{r_{1:T}}, \qquad (2)$$

where $R_{1:T} \equiv \{R_1, \cdots, R_T\}$ and $r_{1:T} \equiv \{r_1, \cdots, r_T\}$ separately represent the CG and AA trajectory over $T$ timesteps, and $P^{CG}(\boldsymbol{R_{1:T}})$ and $P^{AA}(\boldsymbol{r_{1:T}})$ are trajectory ensembles (path probability distributions) of CG and AA atom systems.

In the context of bottom-up CG simulation, AA simulation trajectories are collected as the reference data, which approximate the trajectory ensemble $P^{AA}(\boldsymbol{r_{1:T}})$. The trajectory ensemble of the CG model can also be collected by propagating the system forward in time according to dynamical equations such as Brownian dynamics or Langevin dynamics. To obtain the optimal parameters, one minimizes the discrepancy between two ensembles by minimizing the chosen distance metric between two distributions. For example, using the Kullback–Leibler as the distance metric,[27] one can follow the parameters $\theta$,

$$\theta^{\dagger} = \underset{\theta}{\mathrm{argmax}}\, D_{kl}(Q_{\theta}(R_{1:T}) \parallel P_{ref}(R_{1:T})) \qquad (3)$$

However, optimizing the model parameters $\theta$ to align between the two ensembles according to Eq. (3) is impractical. First, trajectory ensembles reside in a high-dimensional space, making the direct computation of pairwise distance matrices infeasible. Second, the dependence of the trajectory ensemble on the model parameters is highly nonlinear, necessitating specialized optimization algorithms to efficiently explore the parameter space.

To address these challenges, we adopt an adversarial training framework, which provides an efficient solution for aligning trajectory distributions. In particular, the use of a Generative Adversarial Network (GAN) enables the model to learn the underlying distribution of high-dimensional data without explicitly computing high-dimensional distances.[22] The conventional adversarial approach consists of a generator, which produces synthetic samples, and a discriminator, which distinguishes between the generated and target ensemble samples. Through this adversarial learning process, the generator iteratively improves its ability to produce statistically consistent samples, while the discriminator refines its ability to detect discrepancies.

In the dynamics matching framework, we still employ a neural network as the classifier to detect discrepancies. The generator, however, is a CG model that can generate sample trajectories by propagating the state of the system forward in time according to the predefined dynamic equations. In the numerical experiments, we consider two types of dynamics: Brownian dynamics and Langevin dynamics.

In the case of Brownian dynamics, learnable parameters $\boldsymbol{\theta} = \{\boldsymbol{\theta}_U, \alpha, \gamma\}$ include potential network parameters $\theta_u$ for potential $U(\boldsymbol{R}, \boldsymbol{\theta}_u)$, a time scaling factor $\alpha$, and the diffusion coefficient $\gamma$, such that

$$\frac{d\boldsymbol{R}}{d\tau} = -\frac{1}{\gamma}\nabla U(\boldsymbol{R}) + \sqrt{\frac{2k_B T}{\gamma}}\xi(\tau) \tag{4}$$

$$\tau = \alpha t$$

Here for simplicity, we assume a constant diffusion coefficient. Similarly for Langevin dynamics, with the extra learning parameter $m$ (note it need not be the real mass):

$$m\frac{d^2 R}{d\tau^2} = -\nabla U(\boldsymbol{R}) - \frac{d\boldsymbol{R}}{\gamma d\tau} + \sqrt{2k_B T \gamma}\xi(\tau) \tag{5}$$

$$\tau = \alpha t$$

Following the adversarial training framework, we optimize the learnable parameters in the dynamic equation by formulating the problem as a minimax optimization task:

$$\theta^\dagger = \underset{\theta}{\mathrm{argmax}}\left[\min_\phi\left\{\langle f_\phi(\boldsymbol{R}_{1:T})\rangle_{Q_\theta} - \langle f_\phi(\mathcal{M}(\boldsymbol{r}_{1:T}))\rangle_{P_{ref}}\right\}\right], \tag{6}$$

where $f_\phi$ represents the neural network and the expectations $\langle\cdot\rangle$ are taken separately over the CG trajectory ensemble $Q_\theta$ and the AA trajectory ensemble $P_{ref}$. As illustrated in **Fig.1**, trajectories are sampled from both the AA model and CG model. The classifier is trained to give a high classification score for the projected trajectories from the AA model and a low classification score for the CG trajectories. Concurrently, the CG model is optimized according to the feedback of the classifier network, maximizing the classification score of the CG trajectories. To enable this, we leverage JAX-MD[28] as our differentiable molecular dynamics engine, with the classifier network implemented directly in JAX.[29] This framework ensures seamless integration of gradient-based optimization with the dynamics simulations. Other differentiable MD packages, such as TorchMD,[30] offer similar capabilities.

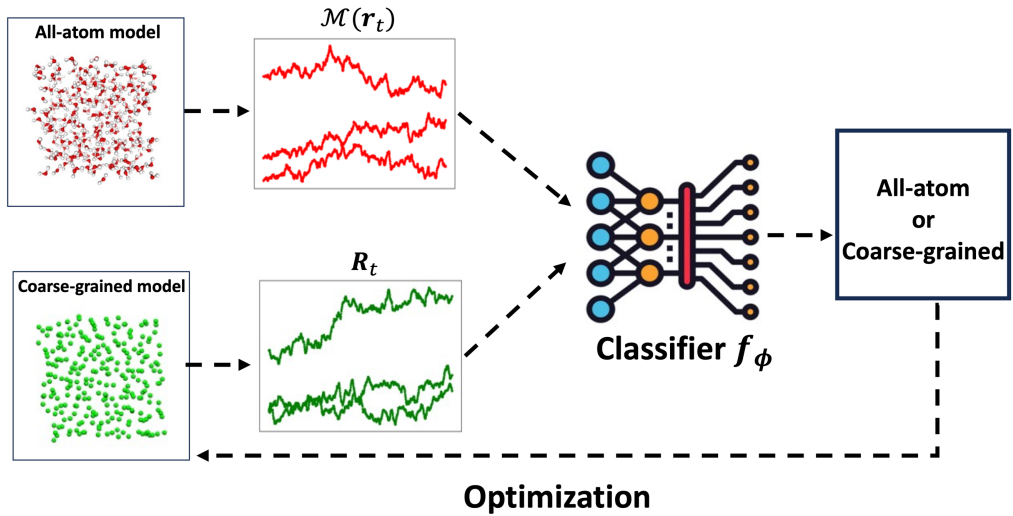

**FIG. 1. Adversarial training framework of dynamic matching.** Trajectory ensembles are separately generated from the all-atom (AA) molecular dynamics and coarse-grained (CG) model. The classifier network $f_\phi$ is trained to discriminate between mapped AA trajectories and CG trajectories. Concurrently, the CG model parameters are optimized to minimize the classifier's ability to distinguish between the two ensembles. Through this adversarial interplay, the dynamic and thermodynamic of the CG model is aligned with the reference AA system.

**Network Architectures:**

In this framework, we are training two networks: the classifier network and the potential network, which characterize the interaction potential between CG beads.

**Classifier network:** The classifier network takes a trajectory $\{R_0, R_1, \cdots, R_T\}$ and predicts whether this trajectory is from AA or CG simulations. Here $R_t \in \mathbb{R}^{3n}$ is the CG coordinate or the mapped atomic representation at time $t$. To ensure the scalability of the network structure, the classifier first uses an equivariant graph neural network (GNN) to encode individual frame $R_t$ into a graph representation $\{v_{k,t}^i, e_{k,t}^{i,j}\}$ where $k$ indexes the GNN layers, superscripts i and j represent the labels of CG beads, v, and e separately represent the node features and the edges features. In the first layer, the directed edge feature from node i to node j is defined by $e_{1,t}^{ij} = R_t^j - R_t^i$, characterizing the displacement between CG beads. The node feature is defined by $v_{1,t}^i = R_{1,t}^i - R_{1,t-1}^i$, characterizing the change of position during the simulations (Fig. 2A). To ensure the effective representation of these features and respect the rotational and translational equivariance property of the system, we adopted the Neural Equivariant Interatomic Potentials (NequIP)[31] network. NequIP is an equivariant graph neural network (GNN) specifically designed to encode atomic environments. It processes node features (atoms) and edge features (interatomic relationships) in a manner that inherently accounts for rotational and translational symmetries. By incorporating these symmetries directly into its architecture, NequIP offers an efficient and physically consistent framework for processing molecular configurations, making it well-suited for encoding molecular representations in simulations.

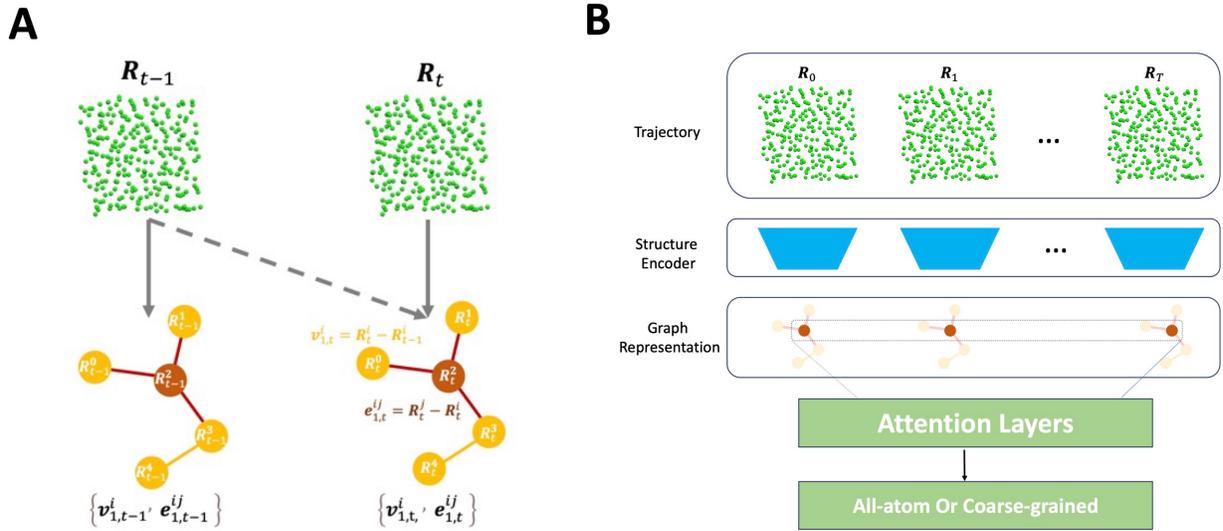

**FIG. 2. Classifier architecture.** (A) Graph representation of particle trajectories. Particle trajectories (represented by Cartesian coordinates) are converted into a graph where nodes correspond to particles, and edges connect nodes if the inter-particle distance is below a predefined cutoff. Node features $\{v_{1,t}^i\}$ encode the displacement of each particle between successive frames, while edge features $\{e_{1,t}^{ij}\}$ represent the relative displacement between connected particle pairs. (B) Classification workflow. The trajectory of graph representations from (A) is processed by an equivariant neural network to update the note features, embedding the structure and dynamic information of the local environment. The time series of updated node features are then analyzed by equivariant attention layers, which aggregate temporal dependencies across frames to compute the final classification score.

Following the encoder, the learned node features are passed into a network that applies the attention mechanism to process the time series information and give the final classification. More specifically, the network takes the time series of individual node features $\{v_{K,1}^i, v_{K,2}^i, \cdots, v_{K,T}^i\}$ as input and gives the classification score for node $i$, denoded as $c^i$. The attention layers in this architecture are analogous to standard multi-head attention layers but adapted to incorporate the equivariant properties of tensor representations. The nonlinear transformations are performed using the e3nn framework [31, 32], which ensures equivariance under rotation. The attention weights are calculated via the inner product between the irreducible spherical harmonic representations of the input features.

The classification score of the whole trajectory is calculated by averaging over all CG beads, i.e. $c = \sum_{i=1}^{n} c_i$ for a system with $n$ beads. This design respects the permutation-invariant nature of the task, ensuring that the classification is independent of the order in which beads are indexed. Additionally, it improves the scalability of the system by reducing computational complexity. The permutational invariance guarantees that the model can generalize to systems of varying sizes and configurations without requiring retraining.

This strategy of using the time series of nodes instead of the time series of graphs to classify the trajectory is based on the assumption that the local environment mostly determines the local dynamics and thermodynamics. The message-passing operation could be considered as an operation that gathers the local information of each node; or in other words, the system is automatically divided into segments, and each node learns a representation that best summarizes the information of each segment for the downstream tasks.

In our experience, for simplicity, the learning process is typically applied to systems with fixed sizes and timesteps. However, the design of the classifier model ensures its adaptability, allowing it to be trained and applied to systems of varying sizes (e.g., different numbers of water molecules) and different timesteps or even various time resolutions, accommodating data collected from diverse simulations.

**Interaction potential network:** We directly employed NequIP architecture as the potential network. Unlike the classifier network, the graph representation in the potential network encodes only the relative displacements between CG beads in the edge features, while the node features represent the types of CG beads. In our specific case of a one-bead model for water molecules, all CG beads belong to the same type, resulting in a constant node feature in the first layer of the GNN. To prevent CG beads from coming excessively close to one another, a repulsive potential of the form $(r_0/r)^2 S(r)$ was incorporated into the network, ensuring physical plausibility and stability in the simulations. Here $r$ is the pairwise distance, $r_0 = 2$ Å, and $S(r)$ is the cutoff function with the form[33]:

$$S(r) = \begin{cases} 1, & r < r_{on} \\ \dfrac{(r_{cut}^2 - r^2)^2 (r_{cut}^2 + 2r^2 - 3r_{on}^2)^2}{(r_{cut}^2 - r_{on}^2)^3}, & r_{on} < r < r_{cut} \\ 0, & r > r_{cut}, \end{cases} \quad (7)$$

where $r_{on} = 2.2$ Å and $r_{cut} = 2.4$ Å.

**Training Details:**

**Stabilizing the adversarial training.** The training process for dynamic matching methods faces significant instabilities, primarily due to two factors. First, the adversarial training framework, while powerful, is inherently unstable because the generator and discriminator are optimized simultaneously with competing objectives. This often leads to oscillatory behavior or divergence in the optimization process. To mitigate this, we introduced a self-supervised discriminator [34] that leverages auxiliary tasks to provide more robust and stable gradient signals. In our case, given the embedded nodes, with a fraction of them being randomly masked, a predictive network is optimized to predict the two properties of the system: (1) The relative positions of CG beads to their neighbors (i.e., the edge feature) and (2) the position shifts the masked nodes. With this self-supervised task, the learned features are forced to contain both structure information and dynamic information. In other words, the network learns both the thermodynamic and dynamical principles of both systems, instead of focusing on a single aspect, and uses that to determine whether the input trajectories are from the AA or CG simulations.

Second, the dynamics of the model are highly sensitive to variations in the parameters $\theta$, resulting in large gradients that destabilize training. To ameliorate the unstable gradient issue, we employed partial backpropagation,[35, 36] a technique that prevents overly aggressive gradient updates by truncating backpropagation in a defined number of steps, thereby maintaining a more stable optimization process. Empirically, we observed that the gradients of potential parameters increased dramatically as the timestep increased, while the gradient of other parameters remained within a reasonable range. This aggressive update of the potential parameters often resulted in excessively large forces or unphysical interactions, ultimately destabilizing the simulation process. Therefore, we employed differential learning rates, assigning a larger learning rate to parameters such as mass and friction coefficients to accelerate the training process, while applying a smaller learning rate to potential parameters to prevent instability. Furthermore, the gradients of potential parameters were exclusively calculated using partial backpropagation to ensure stability. Together, these strategies stabilize training and enable reliable convergence.

### III. Results

To illustrate our method, we investigate this method by building a CG model of liquid water molecules. More specifically, we applied the dynamic matching method to learn from the AA simulation of water. In our experience, the simulation is performed with SPC/E water model[37] within a 2 nm box with periodic boundary conditions. The Langevin integration was used to simulate the constant NPT ensemble with temperature 300 K and pressure 1 bar. An integration step of 2 fs was used, and the structures were saved every 10 steps. The equilibrium run was performed for 100 ns following the energy minimization. A 100 ns trajectory was collected as the data that the dynamic matching method will learn from to build the CG model. The CG map was defined as the center of mass (COM) of each water molecule.

The equilibrium properties we focused on were the radial distribution function (RDF) and angular distribution function (ADF) functions. The RDF is defined as:

$$g(r) = \frac{N}{V^2} \langle \sum_I \sum_{J \neq I} \delta(r - r_{IJ}) \rangle \ , \tag{8}$$

where $r_{IJ} = \|\mathbf{R}^I - \mathbf{R}^J\|$ is the pairwise distance, $N$ is the number of particles, $V$ is the volume of the simulation box, and I and J are indices of particles. The ADF is defined as:

$$p(\theta) = \frac{1}{W} \langle \sum_I \sum_{J \neq I} \sum_{K > J} \delta(\theta - \theta_{IJK}) \rangle \ , \tag{9}$$

where $W$ is a normalization factor to ensure $p(\theta)$ is normalized and $\theta_{IJK}$ represents the angle form angle formed by particles $I$, $J$, and $K$ which are neighbours. A cutoff of 4.5 Å is used to determine the neighbor list. Compared with the RDF, the ADF provides additional structural information by capturing the three-body correlation of neighboring particles.

**Figures 3(a)-(b)** (Brownian dynamics) and **4(a)-(b)** (Langevin dynamics) compare the thermodynamic properties of CG models against their mapped atomistic models. Both the RDF and ADF of the randomly initialized CG models exhibit significant deviations from the AA simulation benchmarks. However, after optimization using the dynamic matching framework, all CG models converge consistently to the AA reference data. This demonstrates that the optimized CG models accurately reproduce the structural spatial correlations inherent in the AA systems.

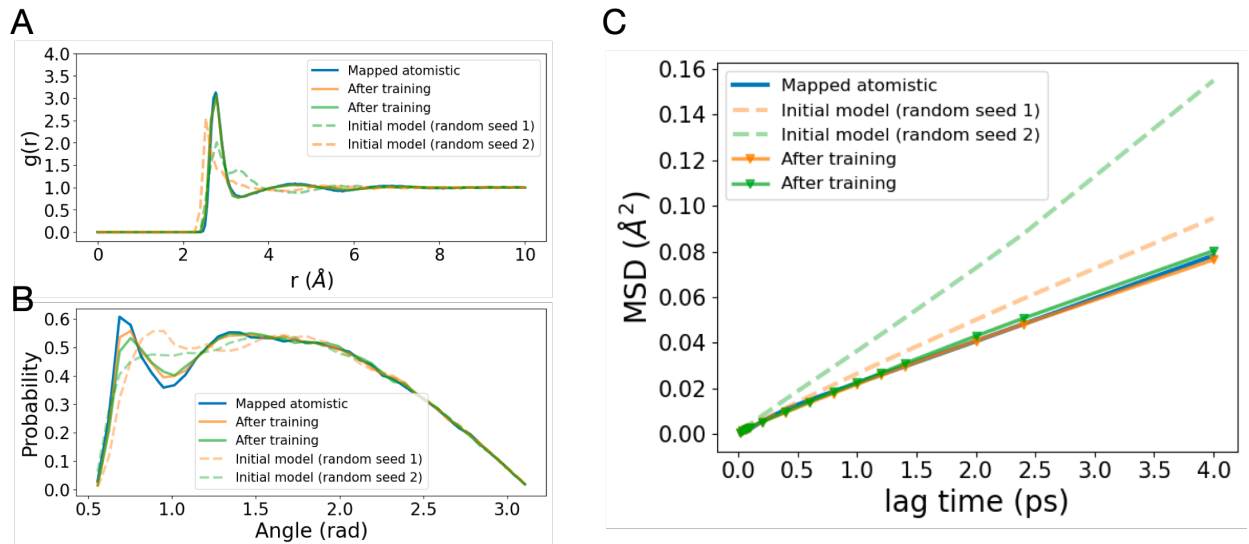

**FIG. 3.** Comparison of coarse-grained (CG) models run with Brownian dynamics and all-atom (AA) simulations. Each panel shows results from two CG models initialized with different parameters, represented in orange and green. (A) Radial distribution function; (B) Three-body correlation; (C) Mean squared displacement. The results from the AA simulations are displayed in blue. Transparent curves indicate results from the initial models, while solid curves represent those from the trained models.

The dynamical property that we focused on was the mean squared displacement (MSD) of the COM of each water molecule, which is defined as:

$$MSD(t) = \langle |\boldsymbol{R}(t) - \boldsymbol{R}(0)|^2 \rangle. \tag{10}$$

The MSD of the CG models is compared against the AA reference data in **Fig. 3(c)** (Brownian dynamics) and **4(c)** (Langevin dynamics). Initially, the MSD profiles of the unoptimized CG models deviate markedly from the AA simulations, reflecting discrepancies in dynamic behavior. After optimization with the dynamic matching framework, the CG models exhibit MSD curves that align closely with the AA benchmarks. This agreement confirms that the CG models not only reproduce structural correlations (as seen in RDFs/ADFs) but also accurately capture the diffusion dynamics of the AA system. Notably, the length of trajectories used for training (0.4 ps) is shorter than the longest lag time shown in **Fig. 3(c)** and **4(c)** (4.0 ps). The training length in fact only goes a relatively small way into the transition to the diffusive regime. This demonstrates that the optimized CG models *can extrapolate essential dynamical features* of the reference system – even when trained on limited short-time data – while retaining predictive accuracy at longer timescales.

For the system studied here, we observed no significant difference in the ability of CG models to reproduce dynamical properties – regardless of whether Brownian or Langevin dynamics was employed. This suggests that, for simple systems like water, the choice of dynamics equation of CG simulations has minimal impact on capturing diffusive behavior. However, we anticipate that the selection of dynamics (e.g., inertial vs. overdamped) may lead to divergent dynamical behavior in CG models of more complex systems, such as polymers, biomolecules, or biomolecular assemblies. A systematic investigation of these effects lies beyond the scope of this work and will be explored in future studies.

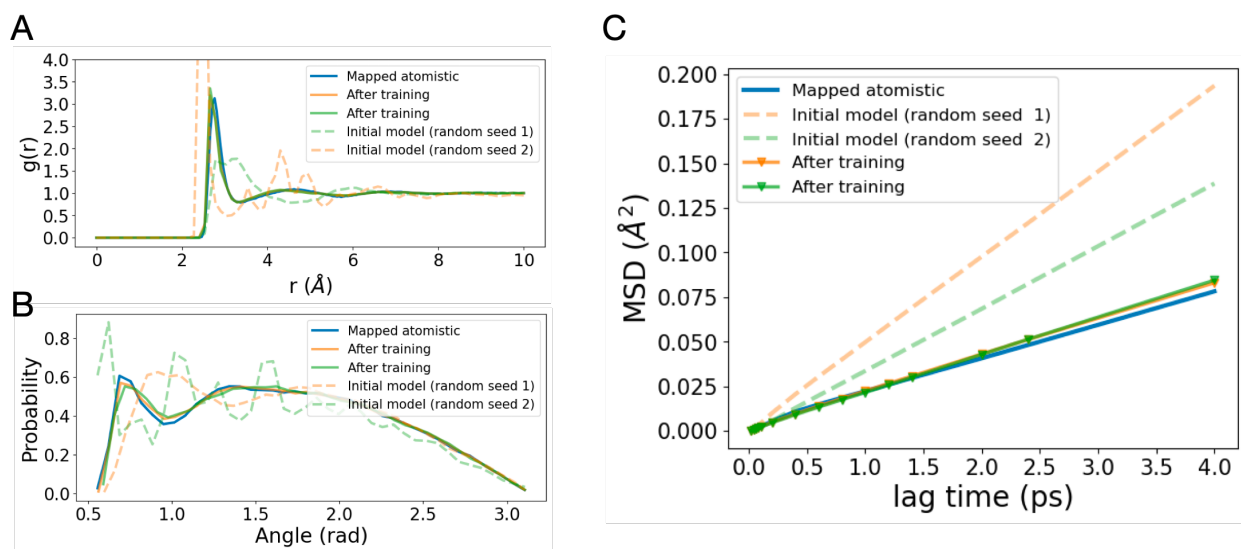

**FIG. 4.** Comparison of coarse-grained (CG) models run with Langevin dynamics and all-atom (AA) simulations. Each panel shows results from two CG models initialized with different parameters, represented in orange and green. (A) Radial distribution function; (B) Three-body correlation; (C) Mean squared displacement. The results from the AA simulations are displayed in blue. Transparent curves indicate results from the initial models, while solid curves represent those from the trained models.

We also evaluated the effect of increasing the CG integration timestep on simulation stability and dynamic fidelity. **Figure 5** presents the MSD of the CG model with different timestep choices. The CG model was trained using a timestep 2.5 times that of the all-atom (AA) simulation. Notably, the MSD curve of the CG model remains close to the AA reference even when the timestep is increased to 5 times the AA timestep. However, we observed that further increasing the timestep leads to unstable simulations. Nevertheless, these results suggest that we can improve the computational efficiency of the CG model by increasing the CG timestep while still maintaining dynamical accuracy. By identifying the largest stable timestep, our method enables longer simulations with reduced computational cost, making it a valuable tool for studying largescale biomolecular systems.

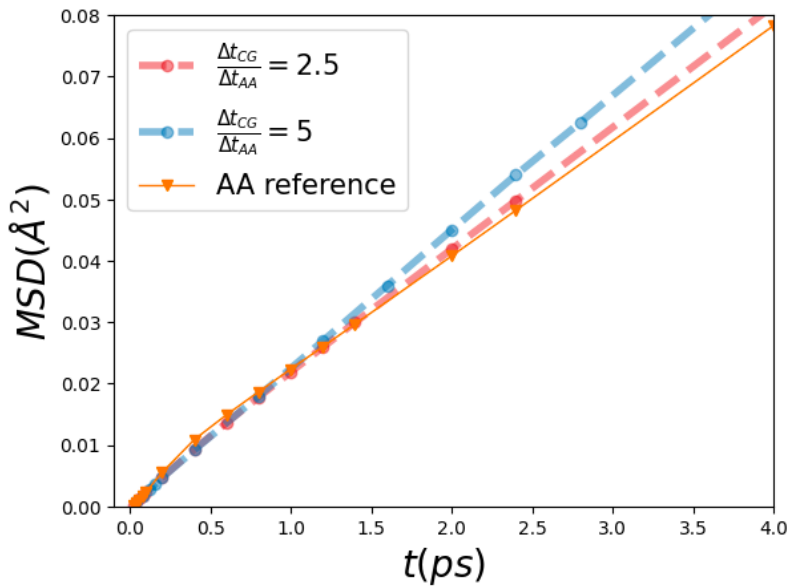

**FIG. 5.** The diffusion behavior of the coarse-grained CG model with varying simulation time steps. The all-atom (AA) reference, performed with time steps $\Delta t_{AA} = 2\ fs$ and saved every 10 steps for training data, is shown as the orange solid line representing the mean squared displacement (MSD) of the center of mass of individual water molecules. The MSD of CG beads is shown for the CG model simulated with time steps $\Delta t_{CG} = 5\ fs$ and $\Delta t_{CG} = 10\ fs$, , represented by red and blue dashed lines, respectively.

**IV. Discussion**

In this work, we introduced a novel adversarial training framework for achieving dynamical consistency in CG models. Unlike traditional approaches that require the preselection of dynamic features, our method aims to match trajectory ensembles directly, ensuring that the CG model faithfully reproduces the underlying dynamics of the AA model. With liquid water molecules as an example, we demonstrated that the proposed method successfully recovers key structural and dynamical properties, including radial distribution functions, three-body angular distributions, and diffusion behavior. Notably, the optimized CG models extrapolate long-timescale diffusion behavior even when trained on short trajectories, demonstrating good generalization. We also showed that our approach allows for robust convergence of the learned parameters across different training runs.

Compared to machine learning emulators, which often struggle with generalization and stability outside their training regimes, the CG modeling framework offers distinct advantages rooted in physics-based principles. By explicitly integrating the governing equations of motion, CG models inherently preserve the dynamical and thermodynamical constraints of the system. While our methodology employs ML components, the temporal evolution of the system remains rigorously governed by these foundational physical laws. This integration ensures that critical mechanistic behaviors are retained, potentially enabling enhanced generalizability across diverse simulation conditions compared to purely data-driven ML approaches.

The design of network architectures and the training framework herein suggest its potential applicability to more complex molecular systems. By employing a graph-equivariant neural network architecture augmented with attention mechanisms for temporal modeling, our method is designed to efficiently process heterogeneous molecular configurations while preserving essential physical symmetries. Although the current work focuses on water as a foundational test case, the architecture's emphasis on scalability lends theoretical support for its adaptability to more complicated systems such as biomolecular assemblies. Future studies will extend this framework to systems where long-timescale dynamics and structural heterogeneity are critical.

While our numerical example adopts Markovian dynamics with a position-independent friction coefficient, the proposed adversarial framework does not impose restrictions on the form of the equation of motion, making it inherently adaptable to a wide range of dynamic models.[38, 39] For instance, as noted earlier it can be extended to incorporate position-dependent diffusion coefficients, enabling the modeling of systems with heterogeneous environments. By leveraging trajectory-level optimization, our method refines traditional bottom-up CG approaches, ensuring that both thermodynamic and dynamic properties are consistently captured.

In conclusion, the adversarial training approach presented here provides a significant step forward in constructing CG models that also preserve the dynamical properties of the underlying atomistic systems while also being more computationally efficient than purely AA MD. By integrating adversarial learning techniques into bottom-up CG'ing, this framework provides a generalizable and physically grounded solution for coarse-graining, thus paving the way for applications to larger and more complex molecular systems. Future work will focus on broadening its applicability to the biomolecular and materials science domains.

## ACKNOWLEDGEMENTS

This material is based upon work supported by the National Science Foundation (NSF grant CHE-2102677). Y.W. gratefully acknowledges the support of the Chicago Center for Theoretical Chemistry Fellowship and the Eric and Wendy Schmidt AI in Science Postdoctoral Fellowship. We thank the University of Chicago Research Computing Center and the National Institutes of Health funded Beagle-3 computer (NIH award 1S10OD028655-01) for computational resources.

## DATA AVAILABILITY

The data that support the findings of this study are available from the corresponding author upon reasonable request.